\documentclass{PoS}
\pdfoutput=1 
\title{The Milky Way's Central Molecular Zone}

\ShortTitle{The Central Molecular Zone}

\author{\speaker{E.A.C. Mills}\\
        Jansky Fellow, University of Arizona Steward Observatory\\
        E-mail: \email{eacmills@email.arizona.edu}}

\abstract{This review compiles the results of recent studies of molecular gas conditions in the central six hundred parsecs of our Galaxy. The review begins by placing our Galactic center into context with the rest of our galaxy. It next discusses the wealth of previous research on the Galactic center, before focusing on what is known about the molecular interstellar medium in this region. It focuses especially on a surge in interest in this region and new studies conducted in the last five years. It concludes by highlighting open questions that remain, and the potential for new facilities such as ALMA to make progress in resolving these uncertainties. }

\FullConference{BASH 2015\\
		18 - 20 October, 2015\\
		The University of Texas at Austin, USA}

\usepackage{graphicx}
\usepackage{natbib}
\usepackage{multirow}
\usepackage{amsmath}
\usepackage{amssymb}
\usepackage{rotating}

\newcommand{\degr}{$^{\circ}$}

\newcommand{\am}{NH$_{3}$}
\newcommand{\hco}{HCO$^+$}
\newcommand{\form}{H$_2$CO}

\newcommand{\cyano}{HC$_{3}$N}

\newcommand{\meth}{CH$_3$OH}

\newcommand{\cm}{cm$^{-3}$}

\newcommand{\msun}{M$_{\odot}$}
\newcommand{\kms}{km s$^{-1}$}

\begin{document}

\section{The Center of the Milky Way Galaxy}

The environment of the Galactic nucleus in the several hundred parsecs surrounding the central supermassive black hole (SMBH) differs significantly from that in the disk in several important ways. First, the average gas density is believed to be $\sim10^4$ cm$^{-3}$, several orders of magnitude above the average in the Galactic disk \citep{Gusten83}. Second, Galactic center molecular gas is extremely turbulent, with line widths of 15-50 \kms\, \citep{Bally87}, in comparison to widths of $\sim$ 1-10 \kms\, typically found in giant molecular clouds in the Galactic disk. The molecular gas is also significantly hotter than gas in the disk, with typical gas temperatures 50$-$100~K \and as high as 400-600~K \citep{Morris83,Gusten85,Huttem93b,RF01,Mills13,Ao13,Ginsburg16}. With its sizable concentration of turbulent, hot, and dense molecular gas, the Galactic center is one of the most extreme environments for star formation in our Galaxy. 

Slightly less than 5\% of the total molecular gas reservoir of $\sim8.4\times10^8$ \msun in the Galaxy is concentrated in the center of the Galaxy, in a region with a diameter of $\sim$ 600 parsecs \citep{Dahmen98, Nakanishi06}. As a result, the surface density of molecular gas in this region is almost two orders of magnitude higher than that typical of the Milky Way as a whole, and is more comparable to gas surface densities observed for ULIRGS and starburst galaxies \citep[see Figure \ref{Fig1}][]{Tacconi08}, high-redshift galaxies, or the centers of more normal, nearby galaxies \citep{MT01,Meier04,Meier08,Kruijssen13b}. While estimates of the star formation rate in this volume vary, they are consistent with being proportional to the amount of molecular gas in this region: $\sim5\%$ of the total star formation rate of the Milky Way as a whole \citep{Chomiuk11,YZ09,Crocker11,Longmore12b,Koepferl15}. However, although the amount of star formation is proportional to the amount of molecular gas, there appears to be a dearth of ongoing star formation, given that the dense gas fraction (the gas with densities greater than 10$^4$ cm$^{-3}$) is believed to be 100\% in this region \citep{Morris89,Beuther12,Longmore13}. As this dense gas should be that which is most relevant to current star formation, and that the majority of the dense gas in the Galaxy is in the Galactic center, one might expect a much higher star formation rate in this region \citep{Lada13}. 

The Milky Way's central concentration of molecular gas thus represents a unique laboratory for studying the mechanisms which control the physical conditions for molecular gas in an extreme environment, and which perhaps ultimately control the star formation process and the degree to which it is universal. Though globally, the molecular gas in our Galaxy represents a quiescent environment for star formation, the molecular gas in the central 600 parsecs is hot, dense, turbulent, and awash in a background of radiation from ultraviolet photons, X-rays, and cosmic rays. The relative proximity of our Galaxy's center allows us to probe gas properties in an extreme environment at spatial resolutions which are possible nowhere else. In this environment it is possible to study the density distribution, kinematics, and heating of the gas, relating them to mechanisms of fueling and feedback, as gas is funneled into the nucleus and as previous generations of stars perhaps shape the formation of new stars. In this review, I present the current picture of the molecular gas in the central 600 parsecs of our Galaxy, focusing on the dynamics of gas clouds in the CMZ (Section \ref{gdyn}), the physical properties of the molecular gas (Section \ref{pcon}), the environment of the Galactic center and its role in heating the gas (Section \ref{henv}), and finally the impact of this environment upon the formation of stars in this region (Section{\ref{sfor}). 
\begin{figure}[tbh]
\vspace{0.1cm}
\hspace{-0.1cm}
\includegraphics[scale=0.4]{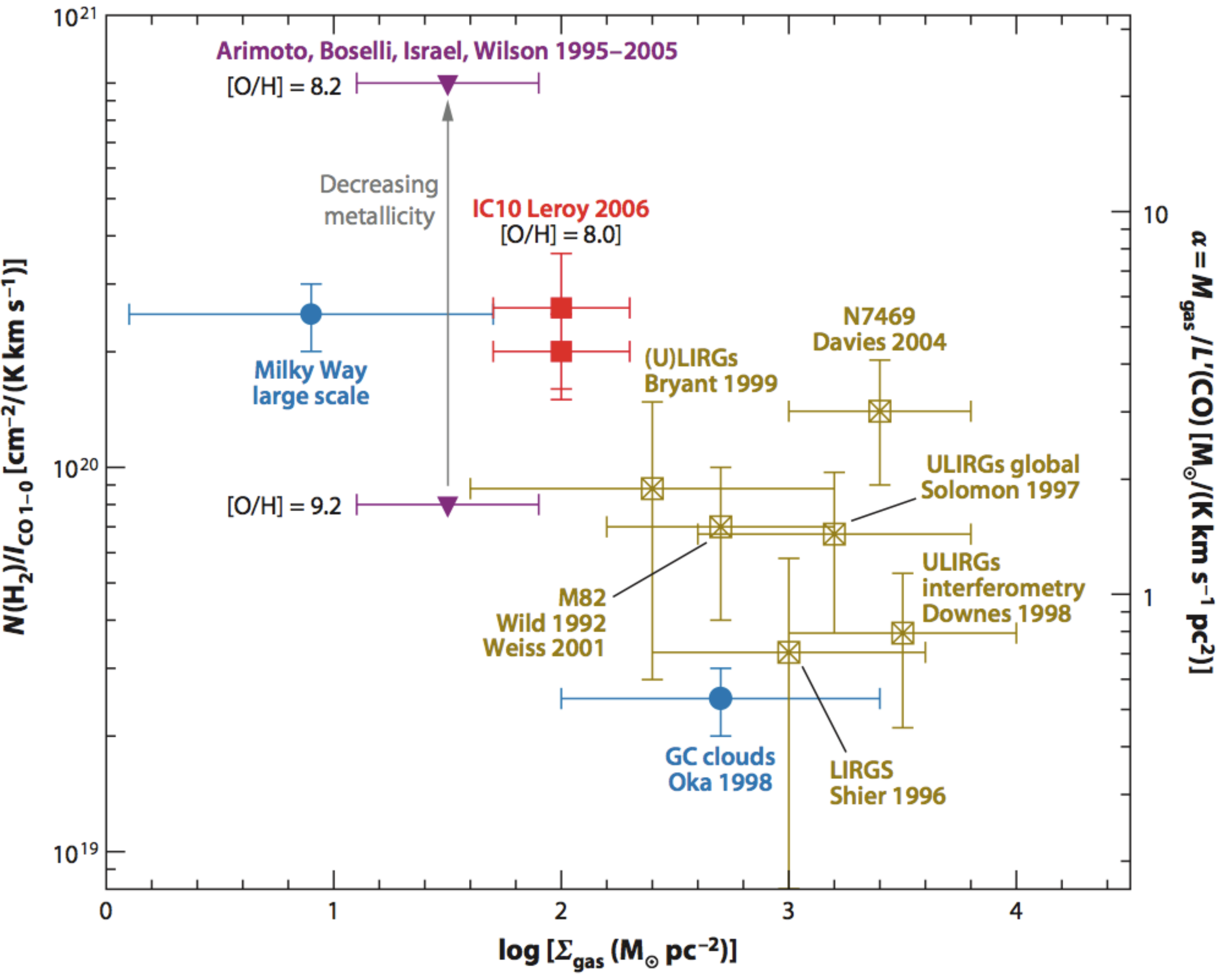}
\caption[Gas surface density in the Galactic center compared to other Galaxies]{ Values of X[CO], the conversion factor from the observed intensity of the J=1-0 CO line to H$_2$ column density. Notice that the surface density of H$_2$ (X-axis) in the center of the Galaxy is almost two orders of magnitude larger than the average for the Milky Way (blue points), and is more similar to typical surface densities in starburst and luminous infrared galaxies (LIRGs), or the centers of more normal, nearby galaxies \citep[e.g., NGC 6946, IC342, and Mafffei 2, which also have X(CO) values 2-4 times lower than the Milky Way][]{MT01,Meier04,Meier08}. This figure is originally from \cite{Tacconi08}; the version here is from \cite{Kennicutt}. }
\vspace{-0.7cm}	
\label{Fig1}
\end{figure}

\section{The Central Molecular Zone}

The Central Molecular Zone, or CMZ, of our Galaxy is defined as the inner $\sim$600 parsecs (or 4 degrees) of the galaxy, where there is a concentration of primarily molecular gas  (Figure \ref{Fig2}). Historically, large-scale surveys of the molecular gas in this region have been primarily carried out using CO and its rare isotopologues \citep{Bania77,LBS77,LB78,Bally87,Bally88,Oka96,Oka98,Oka98b,Oka01,Mizuno04,Oka12}, in conjunction with studies of the neutral gas in HI \citep[e.g.,][]{Burton78,Lang10,McCG12}.  There have also been surveys of molecules which trace a denser gas component-- e.g., HCN \cite{Jackson93}, and CS \cite{Tsuboi99}--  as well as SiO, which traces the distribution of shocked gas \citep{JMP97}. More recently, with the advent of wideband radio and millimeter receivers, there have been simultaneous surveys of dozens of molecular species that better reflect the full chemical complexity of this region \citep{Walsh11,Purcell12,Jones12,Jones13}.  The total mass of molecular gas in the CMZ is believed to be 3$^{+2}_{-1}\times10^7$ \msun \citep{Dahmen98}, with 10\% of this gas concentrated in one giant molecular cloud complex, Sgr B2 \citep{Gordon93}, located $\sim$ 100 parsecs in projection to the east of the dynamical center of the Galaxy. Sgr B2, in addition to being the most actively star forming cloud in the Galactic center, hosts some of the richest known interstellar gas chemistry, detectable in part due to the extremely high column density of this cloud \citep{Snyder94,Miao95,Belloche08}. Outside of the Sgr B2 complex, the remainder of the dense gas in the CMZ is distributed in a population of several dozen giant molecular clouds that are 1-2 orders of magnitude less massive \citep{Longmore13b}. 

\subsection{Gas Dynamics}
\label{gdyn}
     	\subsubsection{Molecular Gas Orbits}
		Individual giant molecular clouds in the CMZ are found primarily on orbits which are controlled by the strong gravitational potential of the nuclear bar \citep{LB80,Binney91}. The first molecular gas is found on the last non-intersecting X$_1$ orbit, which has its major axis aligned with the bar. This orbit crosses the outermost of a separate family of orbits, the X$_2$ orbits, which have their major axes perpendicular to the bar. The resulting large-scale shocks at this orbit intersection along the bar at radii of 200 pc, similar to those observed in other galaxies \citep{Meier,Jogee}, are believed to convert the gas from primarily neutral to primarily molecular \citep{Binney91}, and drive gas inward onto the X$_2$ orbits. At this point, acoustic instabilities in gas at radii where the Galactic rotation curve is still flat will cause gas to continue to flow inward to a radius of 100 pc representing a minimum in the gas shear as the Galactic rotation curve transitions from flat to solid-body, where it is predicted to build up into a ring-like central structure \citep{KK15}. Indeed, the majority of molecular clouds in the central 600 parsecs appear to lie on a `twisted ring' orbit with a radius of $\sim$100 pc and a period of a few $10^6$ years, which is visible in cool dust emission \citep{Molinari11}. Disagreement remains as to the precise structure of this gas: \cite{Kruijssen15} show that closed orbits should not exist in the central potential, and suggest that the gas follows an open orbit, in a `pretzel-like' stream, while \cite{Sofue95} have suggested a model for the gas geometry in which gas on this ring feeds in toward the black hole via several spiral arms. \cite{Henshaw16} also confirm that a closed elliptical orbit for the gas is a poor fit to the kinematic data, but that a spiral arm or open orbit can both reproduce the available data. A nuclear stellar disk with spatial extent and kinematics similar to that of the gas has also been directly detected via stellar spectroscopy \citep{Schonrich15}, which is suggested to be built up from past star formation in the gas on X$_2$ orbits. 
	\subsubsection{Outflows}	
     		Although it is evident that some gas in the CMZ has gone into building up a nuclear stellar disk \citep{Launhardt02,Schonrich15}, it is likely that some of this gas also gets expelled via ionized, atomic, or even molecular outflows, like those observed in nearby starbursts \citep{Walter02,Bolatto13}. Although there is no current evidence for a molecular outflow from the CMZ, there are fossil indications of energetic events that may represent episodes of gas expulsion from the Galactic center, including signatures of a current wind with an entrained neutral component. 
		
		On small scales, there is continuing debate as to whether there is any evidence for a jet from the central supermassive black hole \citep{Falcke00,Markoff07,YZ12,Li13,Rauch16}. Although there are many claimed detections of jet-like structures, this interpretation has been applied to a variety of different features and there is currently no consensus detection of a jet from Sgr A*. On large angular scales, a 1\degr tilted lobe-like structure is observed in the central 100 parsecs of the galaxy, to the north of the Galactic plane. This structure, the northern extent of which was first seen in radio continuum \citep{SH84} and which appears to have a bipolar counterpart in 8 $\mu$m dust emission \citep{BH03}, is suggested to be an outflow from the central parsecs, consistent with being fueled by a starburst \citep{Law08,Law10}.  On far larger scales, gamma-ray observations from the {\it Fermi} telescope uncovered enormous, symmetric lobes extending 50\degr above and below the Galactic plane, which are believed to originate in the Galactic center \citep{SSF10} . Competing theories for their origin involve either jets (or collimated outflows) from an active galactic nucleus (AGN) \citep{ZN12,Guo12,SF12,Yang12}, or the outflow from either a nuclear starburst or a prolonged high rate of star formation in the Galactic center  \citep{Gao12,Crocker12,Carretti13,Crocker15}. A new population of compact, high velocity HI clouds in the central 8\degr$\times$8\degr of the Galaxy may also be related to one or both of these large-scale outflows. More than 80 candidate clouds with velocities up to 200 \kms\, are suggested to have kinematics consistent with entrainment in a starburst-driven outflow \citep{McCG13}. 
		
	\subsubsection{Asymmetries}
		One unusual large-scale feature of molecular gas in the CMZ is its asymmetric distribution. As can be seen in Figure \ref{Fig2}, the highest gas surface densities are found to the east of the dynamical center of the Galaxy, at positive latitudes and velocities. \cite{Launhardt02} estimate that there is three times more mass at positive latitudes as at negative. The gas in the CMZ is also lies in a plane which is tilted with respect to the plane of the Galaxy. One explanation for these asymmetries is that the gravitational potential experienced by gas in the CMZ is asymmetric, leading to an m=1 or 'one-armed spiral' mode: a density wave which orbits around the Galaxy's dynamical center. This is in addition to the m=2 mode which is due to the bar potential. Another possibility is that this instability is a natural manifestation of a long-wavelength  (R$\sim$100-200 pc) acoustic instability in the bar \citep{KK15}. Intriguingly, whatever the dynamical origin of the gas asymmetry, it is also currently reflected in the distribution of massive star clusters-- both of the young massive young star clusters  in the CMZ-- the Arches \citep{Nagata95,Cotera96} and the Quintuplet \citep{Nagata90}, as well as the similarly massive but nascent Sgr B2 star forming region, are found to the east of the dynamical center. This combination of massive clusters and gas at positive latitudes also leads to an asymmetry in the free-free radio emission observed in the Galactic center, highlighting that the eastern side of the CMZ is far more active than the west.

\begin{figure}[tbh]
\vspace{0.1cm}
\hspace{-0.1cm}
\includegraphics[scale=0.4]{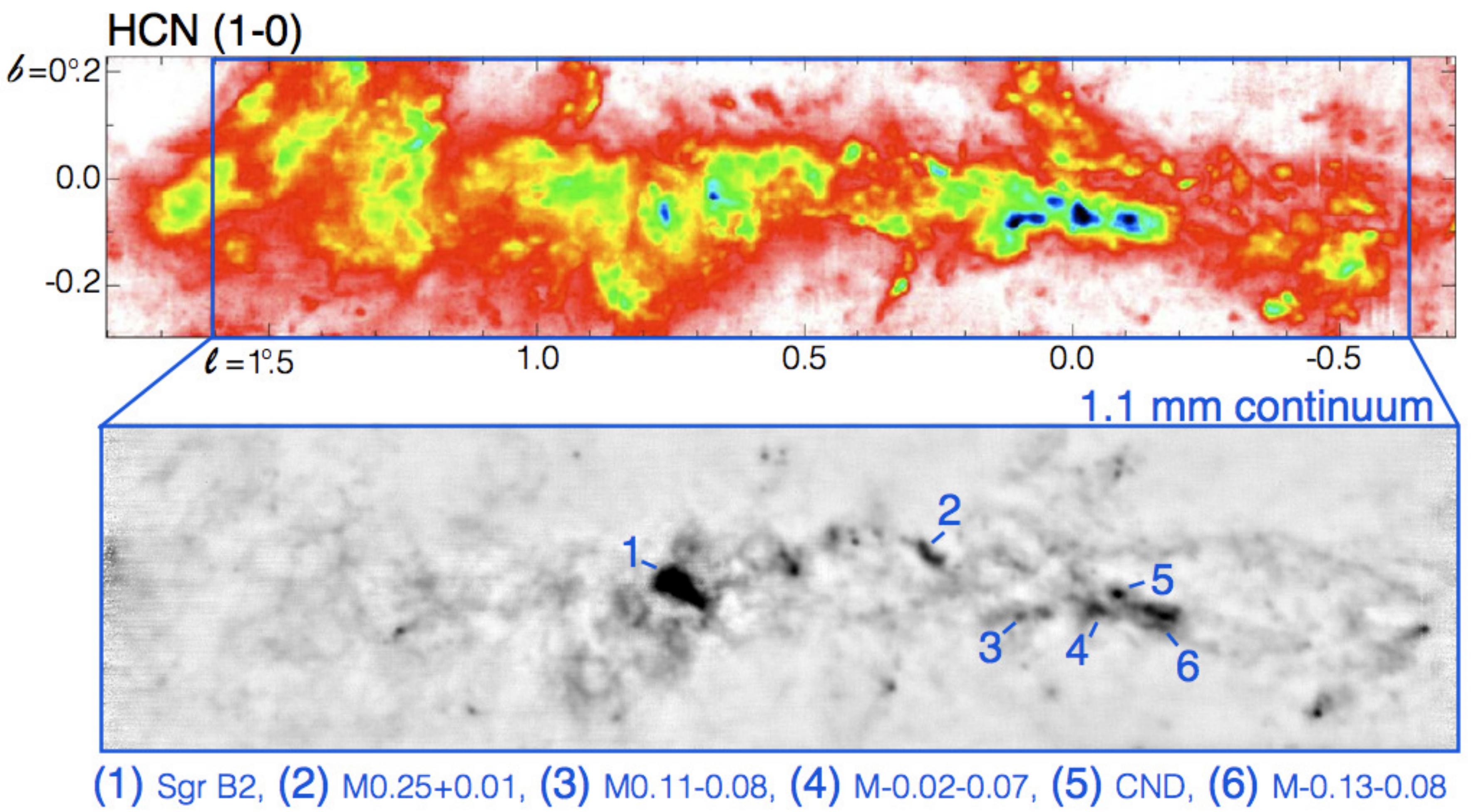}
\caption[The distribution of dense gas in the Central Molecular Zone]{ Top:  Integrated intensity of the HCN (1-0) line, showing the distribution of dense gas in the Central Molecular Zone of the Galaxy. Figure taken from \cite{Jones13}. Bottom: A map of the 1.1 mm continuum emission in the Galactic center \citep{Bally10}, tracing the dense cool dust in the centers of the highest column-density molecular clouds. Individual clouds  are labeled.}
\vspace{-0.7cm}	
\label{Fig2}
\end{figure}

\subsection{Physical Conditions}
\label{pcon}
     	\subsubsection{Density} 
	The highest density gas in the CMZ is found in a population of several dozen giant molecular clouds. Consistent with the overall asymmetry of the CMZ gas, 6 of the 8 clouds with the greatest concentration of mass are found at positive latitudes \citep{Longmore13}.  Global measurements of the average molecular gas densities in the CMZ have been made using several methods. The first is an indirect method, in which it is assumed that observing a large population of molecular clouds indicates that they are relatively long-lived and thus gravitationally bound (which appears at least marginally true for one cloud that has been subjected to a virial analysis). Then, stability at a given distance requires that a cloud possess an average density above some critical threshold-- taking a typical radius of 50 pc from the orbital model of \cite{Kruijssen15}, the required density is $>$10$^4$ cm$^{-3}$ \citep{GD80}, which is generally taken to be the average density of CMZ molecular clouds. 
	
	The simplest direct method is to assume the gas density to be in excess of the critical density of the tracer molecule. In this way, average densities ranging from $\sim10^4$ \cm\, \citep[inferred from observations of CO,][]{Bally88}, $\gtrsim10^4$ \cm\, \citep[from CS 1-0 observations of ][]{Tsuboi99}, and $\gtrsim10^5$ \cm\, \citep[from HCN 1-0 observations by][]{Jackson93} have been found for the molecular gas in the CMZ.  A second direct method is to conduct radiative transfer modeling of non-LTE gas conditions to match observed line intensities to the temperatures, volume densities, and column densities responsible for their excitation. Using lines of CO 3-2 and 1-0, \citep{Nagai07} are able to fit the observed intensities to models with gas densities of $10^{3.5-4.0}$ \cm, though they note they are unable to find solutions to some regions that are affected by self absorption. An excitation analysis using ratios of  3 mm and 7 mm molecular lines, including $^{13}$CS and \cyano\, finds typical densities of a few $10^4$ \cm\, \citep{Jones12,Jones13}. For comparison, \cite{Martin04} performed excitation analyses of CO 7-6 and 4-3, which have higher excitation energies than the lines probed by \cite{Nagai07} or \cite{Jones13} and find densities in cloud interiors up to $10^{4.5}$ \cm, which is the upper limit to which their analysis code is sensitive. Densities in excess of $10^4$ \cm\, are also supported by excitation analyses of H$_2$CO emission \citep{Gusten83,Zylka92}, though apart from these observations of H$_2$CO, there are no large-area excitation analyses of CMZ gas using dense gas tracers. 

	Locally higher densities have been inferred from analyses of individual clouds. Observations of multiple transitions of CS in several clouds (M-0.02-0.07, and the Sickle cloud which abuts the Quintuplet star cluster) indicate that the highest gas densities in these clouds can range from a few $10^5$ from up to a few $10^6$ \cm\, \citep{Serabyn91,Serabyn92}. Similar excitation analyses of multiple transitions of HC$_3$N show that even denser gas can be found in the Sgr B2 cloud, whose mean density is measured to be $10^5$ cm$^{-3}$  and the core of which has densities in excess of 10$^7$ cm$^{-3}$ \citep{Lis91}. Observations of the same molecule in M-0.02-0.07 are fit to two density components, with the lower-density component being several times 10$^3$ cm$^{-3}$  and the high-density component being a few 10$^5$ cm$^{-3}$ \citep{Walmsley86}. In M0.25+0.01, a quiescent molecular cloud which is suggested to be sufficiently massive to form a super-star cluster, densities are estimated to be comparable to the highest densities in M-0.02-0.07, lying between $\sim8\times10^4 $\citep{Longmore} to a few times $10^5$ \cm\, \citep{Kauffmann13}, though these estimates are based on a combination of column density and geometric arguments, and no excitation analysis of the density has been published. 
	
	Apart from Sgr B2, some of the highest densities in the CMZ are suggested to exist in the Circumnuclear disk (CND), a ring of gas and dust surrounding the central SMBH at a projected radius of $\sim$1.5 pc. In the CND, densities up to $10^8$ \cm\, have been inferred from interferometric observations of individual clumps in HCN and \hco, assuming the clumps to be in virial equilibrium \citep{Chris05,MCHH09}. However, single-dish observations with tracers such as dust emission, CO and HCN (molecules for which excitation analyses were conducted), the inferred densities are substantially lower, only a few $10^4$ to $10^6$ \cm\, \citep{Etx11,RT12,Mills13b}. Disagreement over the density of the CND has been suggested to have a substantial impact on our understanding of the nature of the CND, and the future evolution of the central parsecs. The Roche limit for gas at this projected radius from the SMBH is $\sim 10^7$ \cm, which would conventionally determine whether the the gas is dense enough to undergo gravitational collapse to form stars, or whether the clumps that are seen are short-lived structures. However, a more complete virial analysis of the central parsec environment that suggests that the Roche limit might not be physically meaningful in the presence of an external stabilizing pressure \citep{Chen15}. If there is a sufficiently-turbulent interclump medium, then significantly lower densities would not preclude stable clumps or star formation, though the density would still have bearing on the quantity of gas available to form stars and feed the SMBH. 

	Importantly, nearly all of these studies are biased toward measuring the densities in the dense centers of Galactic center molecular clouds. While these structures are often colloquially referred to as clouds, as has been done thus far in this review, their sizes are significantly more compact than typical molecular clouds seen outside of the CMZ: having linear dimensions of typically only 5-10 pc. It is likely that these structures would be better thought of as large clumps or nuclei of a more extended molecular cloud. Consistent with this picture, many of these cloud nuclei have kinematics that would connect them to a larger continuous structure that includes up to a half dozen other cloud nuclei. Examples of larger continuous structures include the `dust ridge' clouds \citep{Lis01,Immer12} and the 50 and 20 km s$^{-1}$ clouds \citep[M-0.02-0.07 and M-0.13-0.08][]{HH05,Minh13}; others can be seen in the `streams' identified in the data used to fit the orbital models of \cite{Kruijssen15} and in the kinematic analysis of \cite{Henshaw16}. These larger structures are likely the true `molecular clouds' of the Galactic center, with the cloud nuclei embedded in a continuous and more diffuse cloud envelope, perhaps one that is being tidally stripped. Thus far, the density structure and extent of these envelopes has not been a major focus of CMZ density studies. However, a very low-density CMZ molecular gas component that could be related to these diffuse cloud envelopes has been detected, first traced by H$_3^+$ \citep{Oka05} in a limited number of sightlines through the CMZ, and followed by observations of numerous additional species with Herschel and other telescopes \citep{Geballe10,Schilke10,Lis10,Lis10b,Sonnen13,Menten11,Monje11,Sonnen13,Lis12}. Most of these species are observed in absorption toward multiple lines of sight against strong infrared and submillimeter continuum sources, especially toward Sgr B2 and Sgr A. The H$_3^+$ observations have characterized this gas component as warm (T$\sim$ 250-350 K), diffuse (n$\sim 10-100$ \cm) and pervasive-- the inferred sizes of the absorbing clouds are several tens of parsecs, and are suggested to constitute a substantial fraction of the volume filling factor in the CMZ \citep{Oka05,Goto08,Goto11}. Conclusively determining the filling factor of this gas and its relation to the more commonly observed cloud nuclei will be important for an accurate determination of the characteristic or average molecular gas density in the CMZ. 
	
	
     	\subsubsection{Temperature} 
     	Just as there is a range of densities measured for the CMZ gas, there is also a wide range of temperatures measured for molecular clouds in the CMZ, from tens to hundreds of K. Hot gas is found throughout the CMZ, from the intersection of the aforementioned X$_1$ and X$_2$ orbits to high-latitude structures suggested to be interacting with the surrounding halo gas \citep{Riq13}. 
		
	The first indication that gas in the CMZ is hotter than gas in the disk of the Galaxy was from \am\, studies toward individual CMZ clouds \citep{Gusten81,Morris83}, indicating average temperatures of $\sim$50 K. Followup studies of larger samples of CMZ clouds using \am\, transitions with energies up to 300 K above the ground state confirmed average gas temperatures of 60-120 K \citep{Gusten85}, and suggested that the temperature distribution could be well approximated by two temperature components, 25 K and 200 K \citep{Huttem93b}. Note that \cite{Huttem93b} find that this cool component is not insubstantial: it is estimated to contain $\sim$75\% of the total column density of \am.  However, despite its suggested dominance, this cool component has been observationally elusive. It is similar to dust temperatures in this region: apart from a few localized enhancements, such as within the central parsec, dust temperatures in CMZ clouds are between $\sim$15-30 K \citep{Mezger86,Lis94,RF04,Molinari11,Etx11}. However, a substantial cool gas component is not detected in \form\, \citep{Ao13,Ginsburg16}, and the only other evidence of such a component is in the lowest-excitation CO lines: an excitation analysis indicates temperatures between 20 and 35 K \citep{Nagai07}, and the observed antenna temperatures of CO (1-0) line are also in this range, which would be equivalent to the kinetic temperature assuming the line to be optically thick and thermalized \citep{Martin04}. The lack of substantial confirmation makes the existence of this component still somewhat uncertain, though no compelling alternatives have been presented: for extremely low gas densities the \am\, temperatures could perhaps be attributed to subthermal excitation, however this would not explain the CO results, which do not depend upon an assumption of local thermodynamic equilibrium. 

	For the higher-temperature gas detected in \am\,(50-200 K), consistent temperatures are also seen in a number of other molecules. \cite{Gusten85} verified temperatures of 50-100 K with millimeter observations of the symmetric tops CH$_3$CN and CH$_3$CCH. Observations of H$_2$CO \citep{Ao13,Ginsburg16}, indicate typical temperatures of 70 K, and up to 150 K (though these observations are not able to accurately measure temperatures if they are above$\sim$ 150 K). Observations of higher-J lines of CO are also consistent with temperatures of 70 K \citep{Martin04}. Observations of the pure-rotational transitions of H$_2$ also indicate a much warmer temperature, 150 K, as well as the existence of an even hotter component (600K), which is also dense \citep{RF01}, though it has only been studied in low column-density environments away from the compact nuclei of CMZ clouds. The 150 K warm gas is estimated to make up 30 \% of the total H$_2$ column, while the hot component contributes less than 1\% to the total column density of gas in CMZ clouds. 
		
	A very hot molecular gas component is also detected in observations of extremely highly excited \am. Temperatures of 250-330 K are seen in 3 CMZ clouds using \am\, lines up to J,K = (7,7) \citep{Mauers86}. Toward Sgr B2, even hotter gas (T = 600-700 K) is inferred from  lines of \am, up to (18,18) having excitation energies up to nearly 3000 K, though this gas is only seen in absorption against the hot core, making its extent unclear \citep{Wilson82,Huttem95,Flower95,Cecca02,Wilson06}. \am\, transitions up to (15,15) have also been detected in emission toward a larger sample of clouds, demonstrating that T$\sim$400 K gas is found apparently universally in CMZ clouds, regardless of whether they are forming stars actively like Sgr B2 \citep{Mills13}. These observations also showed that this hot gas is extended in these clouds on 5-10 pc scales, and that up to $\sim$10\% of the \am\, is found in this high-temperature component. While the most likely explanation is that this is indeed a widespread hot gas component, there is still an alternative to be considered: that this apparently hot gas is actually from a highly-excited population of \am\, molecules that is not thermalized with the rest of this gas, but which instead formed in a highly-excited state. This explanation requires that a significant fraction of the \am\, formed recently, or is in extremely low-density component, in order that it has not undergone sufficient collisions to reach a thermal equilibrium with the rest of the molecular gas. Support for this scenario comes in part from a marked similarity in the rotational temperature components of hot \am\, in Sgr B2 \citep{Wilson06}, and the rotational temperature of `hot'  H$_3$O$^+$ \citep{Lis12}. As H$_3$O$^+$ is a short-lived molecule expected to quickly undergo dissociative recombination upon collision with an electron, it is thus likely to be significantly affected by formation pumping, the signature of which would be an apparently hot component. In the future, temperature measurements sensitive to the existence of this hot gas made using molecules which should not be as sensitive to formation pumping (e.g., tracers that are not symmetric tops, such as H$_2$CO) are needed to conclusively distinguish between truly hot gas and formation pumping. 
	
	\subsubsection{Turbulence}
	Galactic center molecular clouds have elevated turbulent line widths compared to those in the Galactic disk on every size scale that has been probed in this region, from a few tenths of a parsec \cite{Chris05,Kauffmann13,Mills15,Rathborne15} ranging up to tens of parsecs \citep{Bally87,Shetty12}. Such enhanced linewidths are quantitatively similar to those seen in clumps of gas in high-redshift galaxies \citep{Kruijssen13b}. On the smallest scales yet probed, there is a large variation in the linewidths measured for individual subparsec clumps in CMZ clouds, which range from 0.5 \kms\, in observations of N$_2$H$^+$ in a particular quiescent cloud \citep{Kauffmann13}, to 40 \kms\, in the gas that is only a few parsecs from the central supermassive black hole \citep{Chris05}. Typical line full-width half maxima (FWHM) for CMZ clouds on scales of a few parsecs show less variation, and range from 20 to 50 \kms\, \citep{Bally87,Lis98}. Given an isothermal sound speed of 0.3-0.6 \kms\, for gas temperatures of 30-100 K, this implies Mach numbers of 10-60, or highly supersonic turbulence. Although there is a single observation of linewidths less than 1 \kms\, \citep{Kauffmann13}, there is still no firm evidence at the scales that have been probed (down to 0.1 pc) for thermal linewidths in any CMZ cloud (excluding observations of masers whose linewidths are expected to be subthermal). The driving mechanism for the increased turbulence in CMZ clouds has also not been conclusively identified, with candidates including tidal shearing due to differential Galactic rotation \citep{Fleck80,Wilson82,Gusten89} or acoustic gas instabilities \citep{Kruijssen13,KK15}. Additional characterization of the turbulent spectrum over a wider range of length scales is required to identify the driving and dissipation scales of the turbulence. Turbulence also has been suggested recently to play a role both in the suppression of star formation \citep{Kruijssen13} and in setting the initial distribution of masses for star formation \citep{Rathborne14} in the environment of the CMZ. 
	
	\subsubsection{Magnetic Fields} 
	Large magnetic fields were first inferred in the diffuse ISM of the Galactic center from radio observations of nonthermal, linear, and highly polarized filamentary structures stretching over tens of parsecs \citep{YZM86}. Based on the extremely linear structure of these original nonthermal filaments, it was estimated that milliGauss field strengths of were required for their confinement, and suggested that this was indicative of a global, poloidal magnetic field structure in the center of the Galaxy \citep{YZM87,YZM87b,YZM87c}. However, subsequent detection of a larger, faint population of shorter nonthermal filaments, which are randomly oriented throughout this region, has challenged this idea, suggesting that such strong fields are instead localized, and that the overall magnetic field is not highly ordered \citep{LaRosa00,YZ04}.  Additional equipartition arguments have also favored a relatively weak global magnetic field (10 $\mu$G) in the diffuse ISM in this region \citep{LaRosa05}. A comprehensive analysis of these existing magnetic field observations by \cite{Ferriere09}, including additional rotation-measure observations, currently favors the existence of a weak (tens of $\mu$G) global magnetic field, with a generally poloidal geometry. 
	
	In contrast, dust polarization measurements in the submillimeter \citep{Chuss03,Novak03} and near-infrared \citep{Nishiyama09b} show that the field in the dense ISM of the CMZ has an ordered component that is parallel to the Galactic plane. The ordered nature of the field geometry, even in the interiors of turbulent giant molecular clouds, suggests that the magnetic field in these clouds is dynamically important relative to their internal turbulent motions, which is to say that the turbulence is sub-Alfvenic \citep{Pillai15}. Comparing deviations in the magnetic field structure to the turbulent velocities in one cloud (M0.25+0.01), \cite{Pillai15} indirectly estimate a magnetic field strength of 5 mG. There are few direct constraints on the magnetic field strength in CMZ clouds, but a handful of Zeeman observations of several species have been made \citep{SL90,Killeen92,UG95,Marshall95b,Crutcher96,YZ96,Uchida01}. A critical reanalysis of these results has estimated that these observations are consistent with typical magnetic field strengths of a few tens of $\mu$G up to 1 mG in the dense CMZ gas \citep{Ferriere09}, though higher values up to 2-4 mG are detected in the potentially unique environments of the circumnuclear disk in the central few parsecs \citep{Killeen92,Plante95}, and in the OH masers in the Sgr A East supernova remnant \citep{YZ96,YZ99}.  

\subsection{Environment and Heating}
\label{henv}

	As detailed in the previous sections, the diffuse CMZ gas and (at least) a substantial fraction of the dense molecular gas in the CMZ (25-30\%) has temperatures in excess of 100 K, with a smaller, and perhaps more localized fraction of the gas as hot as 600 K. So, what processes are responsible for heating gas in the CMZ to these observed temperatures? Below, I discuss several potential heating mechanisms which may operate more strongly in the Galactic center than in the Galactic disk. 

	\subsubsection{Far-Ultraviolet Background}
		The Galactic center is home to a large population of massive stars, including three massive, young star clusters \citep{Nagata90,Nagata95,Cotera96,Figer99,Figer02} which are ionizing the surrounding ISM on scales of tens of parsecs \citep{Ekers83,YZ87,Lang97,Lang01,Simpson07}, as well as an equal number of massive stars found in the field \citep{Cotera99,Mauerhan07,Mauerhan10,Mauerhan10b,Mauerhan10c}. The effective temperature of the ionizing radiation in the CMZ is estimated to be $\sim$37,000 K  \citep{RF04,RF05}. Molecular gas in the CMZ will be heated through its interface with this radiation field, forming photodissociated or photon-dominated regions (PDRs). PDR heating is one way to explain the observed gas and dust temperature discrepancy: in external regions of a PDR,  gas temperatures will be on the order of several hundred K, with dust temperatures $< 50$ K \citep{Hollenbach91}. Comparing PDR models of \cite{TH85} to observations of ionic, neutral and molecular species in the CMZ, \cite{RF04} find that photoelectric heating through PDRs is consistent with being responsible for 10-30\% of the observed warm (T$\sim150$ K) H$_2$ column density \citep{RF01,RF04}. However, as noted by \cite{Ao13}, PDR heating is not likely important for the densest gas (for example, the 70 K component they trace with H$_2$CO) because heating does not penetrate to the dense cloud interiors where molecules like H$_2$CO are found. Thus, while PDR heating may be important for heating the diffuse gas traced by H$_3^+$ or H$_2$, it is not likely to be responsible for heating the dense gas at temperatures of 50 -200 K traced with NH$_3$-- \cite{RF04} estimate that if \am\, is found in PDRs it would be destroyed on timescale of 7 years--, H$_2$CO, and CH$_3$CN \citep{Gusten85,Huttem93b, Ao13}. Another mechanism must then be responsible for heating the dense gas, and contributing to the observed discrepancy between the temperature of the dense gas and the dust temperatures in CMZ clouds.

	\subsubsection{X-ray Background}
		X-rays, or heating through X-ray dominated regions (XDRs), are another possible heating source for the CMZ gas, as X-rays penetrate more deeply into cloud interiors than ultraviolet radiation.  The Galactic center is an extended source of X-ray emission \citep{Koyama89}, which is believed to originate in a variety of sources, including a population of discrete point sources \citep{Wang02, Muno03,Muno09},  Fe K-$\alpha$ emission from the surfaces of molecular clouds which is interpreted as an X-ray reflection nebula \citep{Koyama96,Murakami00}, as well as some potential contribution from a diffuse, hot plasma \citep{Koyama96}, though the existence of the latter is controversial. Although \cite{RF04} acknowledge that an XDR could penetrate as much as 10$\times$ deeper into CMZ gas than a PDR, potentially sufficient to heat the entire column of observed warm gas in the CMZ, both they and \cite{Ao13} find that the total X-ray luminosity in the CMZ is three orders of magnitude too low to account for gas temperatures $>$100 K. At present then, X-rays are not a dominant factor in the environment of the Galactic center, which, as might be expected, also does not exhibit strong mid-infrared ionic emission indicative of XDRs or active galactic nucleus (AGN) activity \citep{An13}. However, it is worth noting that, if the Fe K-$\alpha$ emission is an X-ray reflection nebula, it requires a recent (within the last few hundred years) event with an energy generation rate of $10^{41-42}$ erg s$^{-1}$ \citep{Koyama96}. If, as is often suggested \citep{Koyama96,Murakami00,Inui09,Ponti10}, the source of this event was an accretion event on the central SMBH, the required luminosity would be six orders of magnitude greater than its quiescent value \citep{Baganoff01}. Depending on the frequency of these and larger events, this could make X-ray heating from intermittent AGN-like activity more viable.

	\subsubsection{Cosmic Ray Background}
		An even more penetrating heating source for CMZ gas is cosmic rays (highly energetic protons and nuclei). A high Galactic center cosmic ray ionization rate of $\zeta \sim2\times10^{-15}$ s$^{-1}$ was first suggested by \citep{Gusten81} to explain the observed discrepancy between dust and gas temperatures in the CMZ.  (For comparison, the typical interstellar cosmic ray ionization rate $\zeta_0$ is estimated to be $3\times10^{-17}$  s$^{-1}$, although recent observations by \cite{Indriolo12} suggest that it may be an order of magnitude higher). More recently, independent observations of absorption from H$_3^+$ toward multiple sightlines are also found to require $\zeta \sim 10^{-15}-10^{-14}$ s$^{-1}$ \citep{Oka05,Goto08,Goto13}. It is suggested that a similarly high $\zeta$ is also required by observations of TeV emission in the central 600 parsecs \citep{YZ13b}, with values as high as $10^{-13}$ s$^{-1}$ inferred for one particularly highly-irradiated cloud \citep{YZ13c}. However, such high values of $\zeta$ may not be a uniform property of the CMZ. While \cite{Harada15} find that chemical modeling of the circumnuclear disk in the central few parsecs is consistent with a cosmic ray ionization rate $\zeta \sim 10^{-15}$, \cite{vanderTak06} observe H$_3$O$^+$ in the envelope of Sgr B2 and find that its abundance is best fit with models having $\zeta = 1-4\times10^{-16}$, much lower than values inferred toward other clouds in the CMZ. Their observations also indicate that $\zeta$ is a factor of three lower in dense gas compared to diffuse clouds, which they attribute to cosmic ray scattering \citep{PS05}. 
		
		Cosmic ray ionization rates several orders of magnitudes above typical interstellar values are a feasible mechanism for heating gas in the CMZ. First, as previously mentioned, cosmic rays penetrate to high column densities, leading to uniformly high temperatures in cloud interiors. Secondly, as cosmic rays preferentially heat molecular gas through a combination of mechanisms including elastic scattering, rotational and vibrational excitation of H$_2$, dissociation of H$_2$, and chemical heating (ionization) of H$_2$ \citep{Glassgold12}, heating by cosmic rays can lead to the observed discrepancy between dust and gas temperatures in CMZ clouds \citep{Ao13}.  Assuming $\zeta = 3\times10^{-14}$ s$^{-1}$,  a model of cosmic ray heating for a typical CMZ cloud indicates that cosmic rays can effectively raise gas temperatures in excess of dust temperatures, even at the high densities of $10^4-10^5$ \cm\, typical of CMZ clouds \citep{Clark13}. At cloud densities of $10^5$ \cm, their model predicts dust temperatures of 17-30 K and gas temperatures of 50-80 K. With this $\zeta$, \cite{Clark13} find that gas and dust do not become thermally coupled until densities exceed a few $10^6$ \cm. Another implication of high $\zeta$ is that it can lead to strong and variable emission from the FeI K$\alpha$ line \citep{Capelli12,YZ13b,YZ13c}. While enhanced and variable iron K-$\alpha$ emission is observed from CMZ clouds, there is an alternatuve explanation for this emission (an X-ray reflection nebula, see previous Section) that also explains the apparent propagation of the emission at the speed of light. Finally, \cite{Papa10} predict that in cosmic-ray-dominated regions, the {\it minimum} gas temperature should be 80-100 K, while in the Galactic center, many clouds appear to have a temperature component $<$ 50 K \citep{Huttem93b,Nagai07}. Thus, while low-energy cosmic rays are a viable heating source, it is not yet certain that an extremely high cosmic ray ionization rate is the globally dominant source of heating in the CMZ.	

	\subsubsection{Turbulence and Shocks}
		The other likely heating source for CMZ gas is the dissipation of supersonic turbulence in CMZ clouds \citep{GK74,PP09}. This mechanism can also explain the observed discrepancy between gas and dust in the CMZ, providing a means of preferentially heating the molecular gas \citep{Wilson82,JMP97,Ao13}. As shown by \cite{Ao13}, turbulent dissipation is sufficient to heat gas with densities between $10^4-10^5$ \cm\, to observed temperatures of 50-60 K. This is consistent with work by \cite{RF04}, who compare models of PDR heating and heating by J (jump) and C (continuous) shocks, finding that low-velocity C-shocks due to turbulent motion are the most likely candidate for heating the bulk of the CMZ gas. Of course, if the dissipation of turbulence is the dominant source of CMZ heating, there must be an energy injection mechanism which maintains the observed turbulence in a steady state. One suggested mechanism is tidal shearing due to differential Galactic rotation \citep{Fleck80,Wilson82,Gusten89}. \cite{Kruijssen13} analyze a number of potential drivers of the turbulence, ruling out energy sources such as feedback and gas inflow. The most likely driver of the turbulence from their analysis is acoustic gas instabilities \citep{Kruijssen13,KK15}.
		
\subsection{Star Formation}
\label{sfor}
	It is clear that the raw material for star formation in the Galactic center is fundamentally different than that found in the Galactic disk.  On scales of tens of parsecs, CMZ clouds are characterized by high nonthermal linewidths, with observed FWHM on the order of 20-50 km$s^{-1}$ \citep{Bally87,Lis98}, believed to be due to turbulence in the cloud interiors. On these same scales, CMZ clouds have gas temperatures far in excess of the 10-20 K gas temperatures typical in disk clouds, \citep[T=$50 - 300$ K:][]{Morris83, Gusten85, Huttem93b, Mauers86}, and they have relatively high densities (n $> 10^{4}$ cm$^{-3}$), which are necessary if the clouds are to maintain their integrity in the face of the strong tidal shear near the Galactic center \citep{Gusten89,Kruijssen13}. These conditions, particularly the large turbulent velocities, may partly explain why CMZ clouds appear largely quiescent and devoid of star formation at these densities \citep{Longmore13,Kruijssen13}, which is remarkable, as disk molecular clouds, when they reach densities of $\sim 10^4$ cm$^{-3}$, are already observed to be actively forming massive stars. 

\begin{figure}[tbh]
\vspace{0.1cm}
\hspace{-0.1cm}
\includegraphics[scale=0.4]{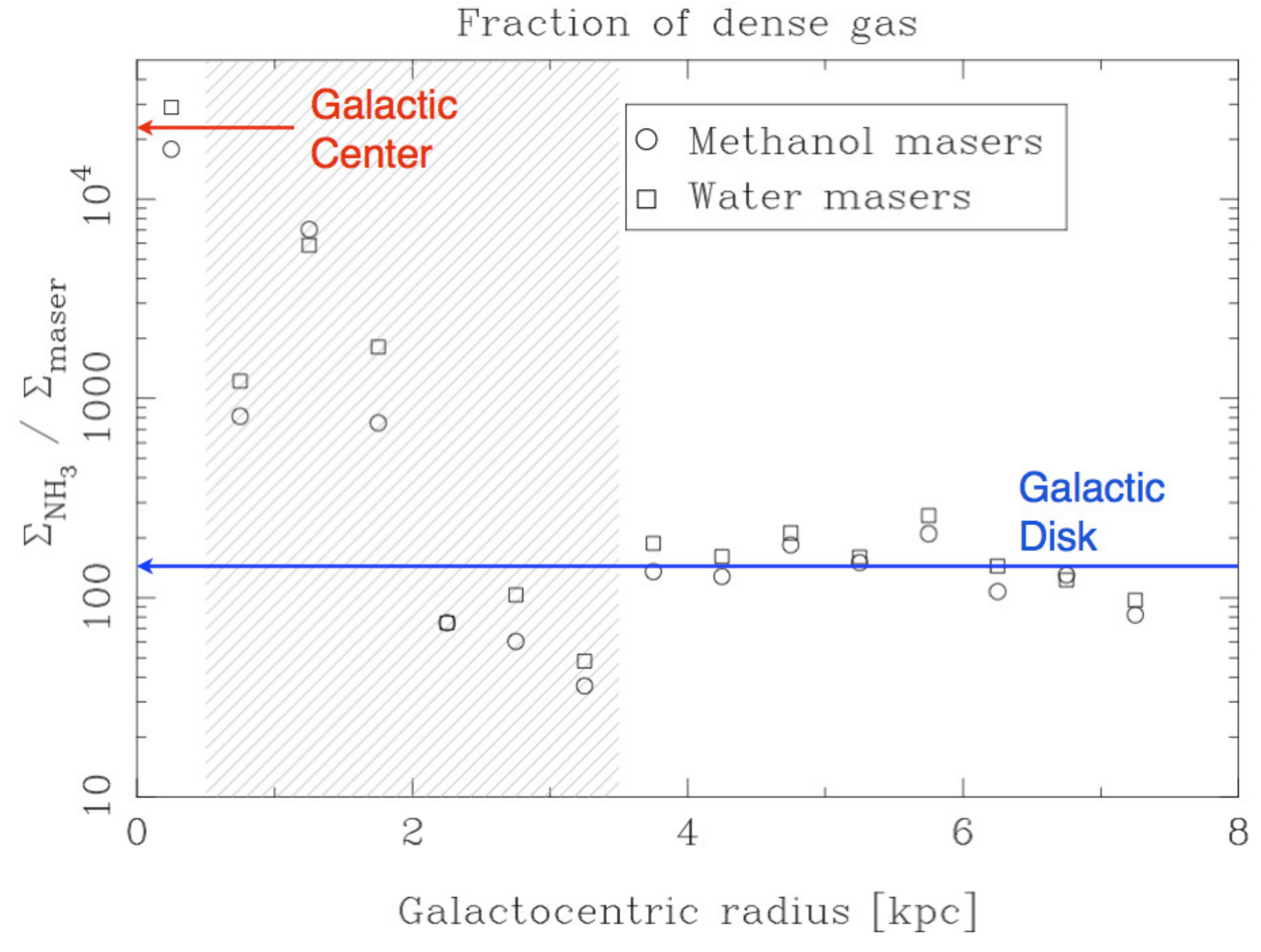}
\caption[Tracers of current star formation per unit gas in the Central Molecular Zone, compared to the Galactic disk]{ A comparison of the amount of dense molecular gas (traced by \am)  associated with each `unit' of star formation (traced by H$_2$O and CH$_3$OH masers) as a function of Galactic radius. The shaded area between 0.5 and 3.5 kpc corresponds to measurements for which where the rotation curve, and thus the derived Galactocentric radii, are not reliable. The Galactic center is observed to have several orders of magnitude more dense gas per associated unit of star formation, compared to typical values in the Galactic disk. }
\vspace{-0.7cm}	
\label{Fig5}
\end{figure}

	But, is the resulting star formation process in the CMZ different than that in the disk? The answer is inconclusive, at least when considering the CMZ as a whole. Based on the oddities of the CMZ environment, including high temperatures, strong turbulence, tidal shearing, and a likely strong magnetic field, \cite{Morris93} predicts that the resulting initial mass function of stars which form in this environment should either favor massive stars, or have an unusually high cutoff mass, below which low mass stars are unable to form.  Examining the products of recent star formation in the CMZ, several groups have found that although the two massive star clusters in this region may have unusual present-day mass functions, this is just as likely to be a function of the evolution of these clusters in the strong potential of the Galactic nucleus as it is to be the result of an initial mass function (IMF) which deviates from a Salpeter form \citep{Espinoza09,Clarkson12,Hussmann12,Habibi13}. The only region where a strong case can currently be made for an unusual IMF is the Nuclear cluster, a large cluster of stars in central parsec. Here, the evidence (including spectroscopic identification of stellar populations and a lack of young, X-ray bright low-mass stars), more clearly supports a flat or `top-heavy' initial mass function \citep{Nayakshin05,Paumard06,Maness07,Lu13}. However, the formation environment for these stars-- in the central parsecs where tidal shear from the central SMBH is extremely strong-- is not likely typical of stars in the CMZ as a whole.  One recent approach, which may help constrain initial mass function in the CMZ as a whole, is to study the so-called `clump mass function' (CMF), the distribution of masses in the substructure of molecular clouds before star formation occurs. Studies have suggested that the shape of CMF is related to the shape of the IMF, modulo an approximately constant efficiency factor \citep[e.g.,][]{Alves07,Lada08,Enoch08}. The first study of the CMF in a CMZ cloud suggests that there may be a flatter clump mass function in regions of the cloud which are being compressed by a nearby supernova remnant, compared to the undisturbed core of the cloud, but the number of clumps studied is still too small for this to be a statistically significant result \citep{Tsuboi12}.
		
	One apparently global oddity of the CMZ is the quantity of dense molecular gas which does not apparently have associated star formation activity, as traced by radiatively-excited \meth\, masers, H$_2$O masers, and young stellar objects \citep[][Figure \ref{Fig5}]{Beuther12,Longmore13, Kruijssen13}.  A case in point is the M0.25+0.01 cloud, which has a mass of several $10^5$ \msun, but no indications of associated star formation apart from a single H$_2$O maser \citep{Lis94,Lis98,Lis01,Longmore,Kauffmann13,Mills15}. This cloud and similar clouds in the CMZ appear to deviate from the observed relationship between gas surface density and star formation (the Kennicutt-Schmidt relation) for dense gas that is observed in other star forming regions and galaxies \citep[e.g.,][]{Lada12}, either suggesting that this relation is not universal \citep[at least not on sub-galactic scales, as is somewhat expected, e.g.][]{Kruijssen14}, or that observational biases are affecting our understanding of Galactic center star formation. For example, surveys are only just beginning to identify protostellar sources in the CMZ \citep[][though note that the latter suggests that some of these may be older stellar sources masquerading as younger protostars]{YZ09,An11,Koepferl15}. There are more sensitive searches for masers which are finding weaker sources indicative of star formation that were missed by previous surveys \citep{Lu15,Ginsburg15}, however it is not clear that this can fully make up for the observed discrepancy. \cite{Kruijssen13} argue that the lack of current star formation in the CMZ is real, and a result of high turbulent pressure in the gas, which raises the density threshold for star formation to $10^8$ \cm\, several orders of magnitude above that typical for the disk ($\sim10^4$ \cm) and which they suggest dominates over other retarding effects such as the magnetic field, radiation pressure, and tidal stripping of clouds. They further put forward a model in which the star formation in the CMZ is is cyclical, governed by the time it takes for infalling gas into the CMZ to build up sufficient densities to overcome the turbulent pressure. This theory has many testable aspects, as it predicts that the strong turbulence in Galactic center clouds extends to small scales within the clouds, and predicts that star formation should only occur in clouds in which the densities are $>10^8$ \cm. 

\section{Open Questions and New Frontiers for Galactic Center Research}
\label{openqs}

Over the past five years, the study of the molecular gas and star formation in the CMZ  has been undergoing a renaissance of renewed interest. This can be traced in part to a confluence of advances enabled by new instruments and observatories. ALMA offers dramatic increases in sensitivity and resolution at wavelengths optimized for the observation of molecular gas and dust, and is allowing for an unprecedented detailed look at the CMZ gas structure \citep{Bally14,Rathborne14,Rathborne15}. In tandem, the newly upgraded VLA affords less extreme increases in sensitivity for observations of spectral lines, but its increase in efficiency for surveying multiple lines in the due to increased bandwidth is also proving revolutionary for surveys of Galactic center gas \citep{YZ13,Mills14,Mills15,Lu15}. The most dramatic increases in wavelength coverage come from the Herschel and SOFIA observatories that have restored access to regions of the spectrum that had been largely inaccessible for decades, providing unique insight into the CMZ gas and dust \citep{Etx11,Molinari11,RT12,Goic13,Sonnen13,Lau13,Lau14,Lau14b,Lau15}. The new abilities of these facilities are also inspiring large-scale investment from older facilities (SMA, CARMA, ATCA)  that do not have same capabilities as ALMA or the VLA in terms of sensitivity or resolution, but can compensate for this by dedicating large chunks of less-competitive observing time to new surveys of these regions \citep{Liu12,Martin12,Kauffmann13,Ginsburg15,Corby15}. Finally, single-dish telescopes like the GBT, Mopra and APEX have been a critical complement to the capabilities of new interferometers with their large-area mapping abilities and sensitivity to extended structure, enabling a new generation of spectral line surveys that point the way for detailed followup by ALMA and the VLA \citep{Jones12,Jones13, Minh13,Ao13, Harada15, Ginsburg16}.

An important role in the increase in attention being paid to this region is also being played by surveys of larger regions of the Galaxy such as ATLASGAL \citep{Schuller09} and the Bolocam Galactic Plane Survey, \citep{Aguirre11} which use improved bolometer technology to efficiently map the dust structure over swaths of the Galactic plane, HOPS which maps \am\, and water masers over the southern Galactic plane \citep{Walsh11}, and Herschel Hi-GAL \citep{Molinari10} and Spitzer MIPSGAL \citep{Carey09} which have uniformly covered the Galactic plane at complementary mid to far-infrared wavelengths. Together, these surveys have thrown the unique properties of Galactic center clouds in sharp relief when compared to exhaustive samples of their Galactic plane brethren \citep{Molinari11,Longmore,Ginsburg12,Tackenberg12}. 

Finally, the increase in interest in the gas and dust in the central hundreds of parsecs of the Galaxy is further being driven by the increased focus (and now, thanks to many of same new facilities, increased observational accessibility) of distant high-redshift galaxies with properties more extreme than those typically seen in the molecular gas in our Galaxy \citep[e.g.,][]{Carilli13}. The CMZ is increasingly seen as a testbed for conditions of extreme star formation like those in high-z objects that may stress or break existing, universal understanding of this process \citep{Bastian10,Lada12,Kruijssen13b}

Despite this renewed interest however, many old questions remain (and even more new questions are being provoked). Further, the wealth of new studies only adds to a surplus of observations of this popular region (observations that often have discontinuous coverage, disparate molecular tracers, mismatched brightness sensitivity, or significantly different sensitivity to spatial scales), the interpretations of which remain difficult to reconcile with a single unified model of the gas in this region. Below, I highlight a few of these questions, discuss how existing data can be better connected to build a unified picture of the CMZ, and note a number of the key observations of this region that are still lacking. Finally, the review ends with an exploration of the ways that current and upcoming facilities will be able to place the CMZ in an unprecedented context with observation of our own and other galaxies.

\subsection{A Selection of Questions}

{\bfseries What is the Physical Structure of CMZ gas?}
	To begin with, one of the major unanswered observational questions is the structure of the CMZ molecular gas, both geometrically (the qualitative distribution and quantitative filling factors of all of the gas kinematically associated with a single `cloud' in this region) but also in the co-varying distribution of physical properties in a cloud: its temperature, density, and turbulence. Individually, many different values of temperature, density, and turbulent linewidths have been measured throughout the CMZ. Taking temperature as an example, these values have been observed to differ on various size scales, between clouds, or with the tracer used to measure them \citep{Ginsburg16}. Generally, these properties are measured in heterogeneous datasets, so it cannot be determined to what degree observed variations are due to systematic or measurement uncertainties, or in the case of such systematic variations, which tracers give reliable measurements. Up until now, analyses have also largely been focused upon constraining each property (e.g., temperature, density, turbulence) independently, and thus in the case of disagreement between properties measured using a given tracer, it is not always clear whether each tracer may be preferentially probing a subset of gas which is different along multiple axes (for example, that the high-temperature gas is also the high-density gas, and thus is accessible with an entirely different set of tracers than the low-temperature, low-density gas). A significant improvement in our understanding of the gas physics will be enabled by connecting the measurements of temperature and density to determine how they co-vary, (as a function of both scale and position) in individual clouds. Further, the models of temperature and density that are fit to the observations have thus far been extremely granular, with division into one, two, or occasionally three discrete gas components. In reality, the gas is likely to consist of a continuum of temperatures and densities, and more accurate fitting of models in the future will need to be sensitive to continuous distributions of these properties.  Accurately determining the structure of the CMZ gas clouds is important for interpreting their disparate properties (some are actively star forming, while most are not) and for detailed testing of models of heating, cooling, and magnetohydrodynamic turbulence in this extreme, high-pressure environments.  

{\bfseries What is the 3D Distribution of CMZ Structures?}
	If the previous question addresses the nature of the fundamental building blocks of the CMZ, this question asks how these building blocks are arranged within the volume, and with respect to other components including the atomic and ionized gas, the stellar disk, massive star clusters, supernova remnants, and the central supermassive black hole. Here, the goal is to overcome our necessarily edge-on view of the CMZ and to reconstruct line of sight positions of all of the observed components, particularly (in the case of the molecular gas) disentangling the local kinematics from the orbital motion. An accurate model of the 3D structure of a multitude of components in the CMZ is required for a number of purposes, beginning with testing and verifying existing orbital models of the molecular gas \citep{Sofue95,Molinari11,Kruijssen15,Henshaw16}. An accurate 3D picture is also needed to determine the dependence of physical properties of the gas with orbital position and distance to the black hole \citep{Ginsburg16}, to determine the current rate of feeding gas toward central supermassive black hole, and to assess any local sources of feedback affecting the properties and star formation potential of CMZ clouds. In combination with the previous question, the determination of the filling factor of the dense gas that such a model will enable will also help in comparing the dense gas fraction and deprojected surface density with expectations from star formation laws \citep{Longmore12b,Lada12}. Finally, determining the 3D location and motions of gas are also needed to put our Galactic center into context by comparing its observed properties and distribution to other (face-on) galactic centers.
	
{\bfseries How does CMZ Gas Evolve With Time?}
	Understanding where and how components of the CMZ are distributed at present would also would bring us closer to answering the related questions of how this environment will evolve in the future, and how it has looked in the past. One of the biggest questions is how much time the Galactic center spends in a steady (low or normal) star forming state, and how much time it might spend in more violent starbursts. In particular, if there are starbursts, how frequent are these bursts, and what ultimately triggers them? One can also ask how long it takes to build up gas (and if this sets typical cloud lifetimes in the CMZ) as well as where the gas for starbursts typically originates: is it all fed smoothly into the CMZ through the bar, or is there any external accretion from, for example, satellite galaxies or streams of halo gas that might trigger nuclear starbursts?  As the most recent orbital models \citep{Kruijssen15} place the bulk of gas $\sim$50 pc from the black hole, it could also be asked how CMZ gas in the course of its evolution is further fed into the central parsecs and beyond. Is AGN activity from this feeding connected to starbursts? There is already some indication of both starburst and AGN activity within the past 10 Myrs in the CMZ, in the existence of the Fermi bubbles \citep{SSF10} and three similarly-aged young massive star clusters \citep{Hussmann12,Clarkson12,Lu13}. Answering all of these questions in the center of our own Galaxy would have important implications for more broadly understanding these phenomena in galaxies in general. 
	
{\bfseries Is the Galactic Center the Same Kind of Extreme as High Redshift Galaxies?}
	Although a quantitative similarity has been noted between some physical conditions in CMZ clouds and high-redshift galaxies, most notably in their turbulence for their size and their resulting pressure \citep{Kruijssen13b}, other properties have not yet been clearly shown to be analogous. Temperatures and gas densities are less well determined for high-redshift systems (and up to now, have generally relied on different tracers than those used to study CMZ gas). Further, some properties are known to be quite different, most notably the metallicity of the gas, which is high in the galactic center compared to the disk, let alone compared to the primordial environment of such distant galaxies. The metallicity might be expected to affect the gas chemistry, heating, and cooling, and thus to perhaps alter the observed gas conditions in the presence of the same underlying physical mechanisms dominating the energetics of the gas. Thus, two questions remain to be fully answered: to what degree are the conditions (temperature, density, turbulence, metallicity, magnetic field) quantitatively the same for regions of the CMZ and large samples of high-redshift galaxies, and are the physical mechanisms responsible for these conditions (star formation feedback, AGN activity, galaxy mergers, disk instabilities, cosmic rays, shearing) the same as well? Determining the limits within which this comparison is valid and the CMZ is an adequate analog for high-redshift systems is necessary if the CMZ is to be used to its full potential for dissecting the detailed, sub-parsec gas properties and physics within extreme environments. 
	

\subsection{Fitting the Puzzle Pieces Together}
	Answering the above questions will require many things: an improved synthesis of existing observations, new observations enabled by new facilities, observations that are not yet possible with current generations of telescopes, and improvement in our current methods of analysis.  For example, a complete picture of the distribution of the components in the CMZ will require a comprehensive analysis of multiwavelength data significantly beyond the copious, complex, and at times self-contradictory observations of just the molecular gas presented here. Even with new observations, determining the true distribution of derived physical properties such as temperature and density will likely necessitate a statistical approach to sift through and appropriately weight the large volume of existing data, assessing its reliability, completeness, and systematic limitations. Below, a few additional examples of new observations and analysis techniques are noted. 

	\subsubsection{What are the key missing observations?}
	There are a number of observations that have not been possible or feasible in the past, but which should be enabled by the capabilities of telescopes such as ALMA and the VLA. One of the most obvious is wider probes of  the magnetic field strength and orientation in CMZ clouds. Previously, the broad linewidths of gas in the CMZ have limited measurements of the Zeeman effect to sensitivity to regions of extremely large field strengths, or to maser sources whose conditions may not be representative of typical CMZ gas. However, the increased sensitivity and spatial resolution of facilities like ALMA should now allow these measurements to be made in smaller structures with correspondingly narrower line widths, thus allowing more sensitive determinations of the field strength. Increased sensitivity will also allow better probes of the chemical richness, not just in the high-column density environment of Sgr B2, but in more typical (and quiescent) CMZ molecular clouds outside of Sgr B2, to better assess the uniqueness of gas chemistry in a galaxy nucleus away from the effects of massive star formation. Such molecular line surveys across a wide range of frequencies will also allow for a needed coupling of measurements of the chemical and excitation conditions to enable a more robust determination of both the chemistry and physical conditions, and their variation in the CMZ gas. Just as these studies will be aided by increased sensitivity that allows for surveying larger contiguous regions of the CMZ, so too will studies of the turbulence, which rely on observations that are sensitive to structure on size scales spanning orders of magnitude in order to reconstruct a turbulence spectrum. 
	
	Finally, one class of observations that is so far quite limited is direct `apples to apples' comparisons between the CMZ and the rest of the Milky Way, and between the Galactic center and other nearby galaxies. There are a number of phenomena-- for example, highly excited ammonia indicative of hot gas \citep{Wilson06,Mills13}, and widespread methanol masers indicative of widespread shocks \citep{YZ13,Mills15}-- which are suggested to be unique to the CMZ, but also have not been systematically searched for in other parts of the Galaxy. Not only are such searches possible with the new VLA, thanks to the new capabilities of ALMA, it will also be possible for the first time to conduct observations matched in both resolution and molecular tracer that directly compare gas in the Galactic center to that elsewhere in the Galaxy on the size scales of individual forming stars and protostars. This work will allow any differences in the star formation process between the extreme, high-pressure environment of the CMZ and the more quiescent environment of low and high-mass star forming regions in the Galactic disk to be quantified on the spatial scales most relevant to star formation in terms of the temperature, density, and turbulent structure of the gas as well as their protostellar populations and clump mass functions or column density probability density functions (PDFs). ALMA will also allow the molecular gas on the size scales of parsecs to be probed in a half dozen nearby (d $\sim$ 3 Mpc) galaxies with a variety of molecular tracers. This will enable a different type of comparison, that of the global gas properties in the CMZ made with single dish surveys (having a comparable characteristic parsec-scale resolution) to gas in nearby galaxies on identical parsec scales that resolve individual molecular clouds. New array receivers on these single-dish telescopes (such as the KFPA on the GBT and the upcoming LASMA on APEX) will continue to revolutionize spectral line mapping of large areas, allowing for yet more of the CMZ to be mapped in a larger number of molecular tracers. This will allow an even more complete analysis of the variation of gas properties with environment, as this sample of galaxies includes those with yet higher star formation rates, more active black holes, and lower metallicities than those found in our CMZ. 

	\subsubsection{Improving our Analyses}
	In addition to new data, significant progress in the future will also rely upon improved analysis techniques. The extremely complicated motions and overlapping distribution of gas in this region has previously frustrated typical kinematic analysis techniques, such moment analysis \citep[e.g.,][]{Jones13}. Improved techniques for automated spectral component fitting in this complex environment to pull out common components shows promise for overcoming many of these difficulties \citep{Henshaw16}, and will be necessary for taking chemical and excitation analyses of multiple spectral lines beyond the analysis of discrete regions and even continuous 2D areas, to directly fitting all of the 3D information in spectral line cubes. Spectral fitting of multiple species simultaneously, breaking them down into regions where the spectral profiles do and do not overlap, will also enable a better determination of where the gas being detected in two different molecular tracers is identical, allowing the physical properties (temperature and density) to be compared and correlated. As previously mentioned, such analyses also stand to be improved by fitting continuous distributions of temperature and density rather than the discrete approach that has exclusively been applied to this environment. Finally, our analyses stand to be significantly improved by going beyond such traditional methods as discrete spectral line fitting, moment maps, PV diagrams, and `movie' displays for the analysis of spectral lien cubes of molecular gas. New facilities have the ability to generate datasets with thousands of channels and hundreds of spectral lines, making the analysis of these datasets with these traditional methods (one can hardly watch let alone compare all such cubes as movies) unfeasibly time consuming or even impossible, without throwing away useful information to compress the data into something more cognitively approachable. Seeking out expertise from the fields of data visualization and computational scientific imaging may provide entirely new solutions to these problems that will allow us to make the most of the upcoming big data revolution in radio and millimeter astronomy. 

The last five years have seen a huge resurgence in studies of the Galactic center, with a host of new observations, models, and findings. The next 5-10 years of observations with ALMA, the VLA and SOFIA should enable even more significant progress in studies of the CMZ. Combined with the mid-IR capabilities of JWST for tracing ionized gas, hot dust, and rovibrational molecular lines, it should be possible to get an entirely new comprehensive and high-resolution multiwavelength view of the CMZ molecular gas that will bring us closer to understanding the physics behind the unique gas properties in this region, and how they connect to the processes at play in the present day in the rest of our galaxy, and those that dominated far in the past in high-redshift galaxies of the early universe. 

\clearpage

\bibliographystyle{JHEP}
\bibliography{review}

\end{document}